\documentclass[12pt]{article}
\usepackage[margin=1in]{geometry}
\usepackage{amsmath,amssymb,amsthm,mathtools}
\usepackage{lmodern}
\usepackage[T1]{fontenc}
\usepackage{setspace}
\usepackage{float}
\usepackage{capt-of}
\usepackage{graphicx}
\usepackage{needspace}
\usepackage{placeins}
\usepackage{flafter}
\usepackage{booktabs}
\usepackage{enumitem}
\usepackage[dvipsnames]{xcolor}
\usepackage[authoryear,round]{natbib}
\usepackage[colorlinks=true,linkcolor=blue!50!black,citecolor=blue!50!black,urlcolor=blue!50!black]{hyperref}
\hypersetup{pdftitle={The Welfare Limits of Market Segmentation with Captive Consumers},pdfauthor={Irfan Tekdir},pdfkeywords={Price discrimination, Market segmentation, Captive consumers, Information design, Price dispersion}}
\theoremstyle{plain}
\newtheorem{theorem}{Theorem}
\newtheorem{proposition}{Proposition}
\newtheorem{lemma}{Lemma}

\theoremstyle{definition}
\newtheorem{assumption}{Assumption}
\theoremstyle{remark}

\usepackage{etoolbox}
\newlength{\stmtparindent}
\newcommand{\keeptogether}[1]{%
  \BeforeBeginEnvironment{#1}{\par\addvspace{\medskipamount}\noindent
    \setlength{\stmtparindent}{\parindent}\begin{minipage}{\linewidth}\setlength{\parindent}{\stmtparindent}}%
  \AfterEndEnvironment{#1}{\end{minipage}\par\addvspace{\medskipamount}}}
\keeptogether{theorem}\keeptogether{proposition}\keeptogether{lemma}\keeptogether{corollary}
\keeptogether{definition}\keeptogether{assumption}\keeptogether{remark}\keeptogether{example}

\newcommand{\ps}{{\pi^*}}
\newcommand{\E}{\mathbb{E}}
\newcommand{\Csym}{C^{\mathrm{sym}}}
\newcommand{\U}{\mathcal{U}}
\newcommand{\Q}{\mathcal{Q}}
\newcommand{\atanh}{\operatorname{artanh}}

\title{The Welfare Limits of Market Segmentation\\ with Captive Consumers\thanks{I used Anthropic's Claude as a research assistant for editing, typesetting, and checking derivations and references. All errors are my own.}}
\author{Irfan Tekdir\thanks{Department of Economics, The Pennsylvania State University. Email: \href{mailto:tekdirfan@psu.edu}{tekdirfan@psu.edu}.}}
\date{October 28, 2026}

\begin{document}
\maketitle

\begin{abstract}
\setstretch{1.05}
\noindent In a duopoly with captive and contested consumers, I characterize every pair of consumer and producer surplus that can arise under some market composition and some public segmentation, and construct a market that produces each pair. The set is lens-shaped. At any level of industry profit, consumers do worst when captive and contested consumers are perfectly separated and best in an unsegmented market in which the two firms have equally many captives. A market with unequal captive bases and some contested consumers cannot reach this upper frontier under any segmentation. Between Bertrand competition and monopoly, the frontier itself lies strictly below what consumers would get if everyone paid a single deterministic price yielding the same profit. Firms must randomize, and with downward-sloping demand the resulting price dispersion lowers consumer surplus.

\vspace{\baselineskip}
\end{abstract}

\section{Introduction}\label{sec:intro}

Some consumers compare offers. Others buy from one firm out of habit, loyalty, or inattention: a bank's long-standing depositors, a retailer's app users, households that have never switched energy supplier. Firms can now use data to tell these two groups apart, and data brokers, platforms, and regulators decide how much of this information reaches them. This paper asks which combinations of consumer surplus and industry profit this information can produce when one does not know how many consumers are captive to each firm. The natural place to start is the case in which that number is known.

\citet{armstrong2019} study that case in a duopoly with captive and contested consumers, a model that goes back to the sales models of \citet{varian1980} and \citet{narasimhan1988}. They compare uniform pricing with \emph{discrimination}, under which firms can tell captive from contested customers. Each firm then charges its captives the monopoly price, and Bertrand competition drives the price paid by contested consumers down to cost. Discrimination harms consumers when the firms are nearly symmetric. When the firms are sufficiently asymmetric and demand is not too convex (log-concave demand is enough), it benefits them. The step that makes their analysis work, and that I use throughout, is to write a consumer's surplus as a function of the profit she generates. Under a mild condition this function is concave, so consumers in aggregate are averse to variation in profit.

\citet{csz2026} place the same model in the information-design framework of \citet{bbm2015}. Instead of comparing two regimes, they ask which public segmentation of a given market is best or worst for producers, consumers, and total surplus. With downward-sloping demand the problem is not zero-sum, because high prices reduce the quantity sold as well as transferring surplus. They show that it is enough to consider symmetric segments and nested segments, in which only one firm has captives. The producer-optimal segmentation is nested and always lowers consumer surplus relative to uniform pricing, while the consumer-optimal segmentation combines a symmetric and a nested segment and can be a Pareto improvement.

Both papers take the composition of the market as given, meaning the share of consumers captive to each firm. Outside parties rarely know it. A competition authority assessing a data-sharing arrangement, or a platform deciding what to reveal to sellers, usually knows more about demand than about the mix of captive and contested customers, and that mix differs across submarkets and changes over time. So I ask a question that does not require knowing the composition. Holding demand fixed, which pairs of consumer and producer surplus can arise under some market composition and some public segmentation? The answer bounds what any information policy can achieve, whatever the composition, and the construction tells me which compositions and segmentations produce the best and worst outcomes for consumers.

\paragraph{Results} The achievable outcomes form a lens (Theorem~\ref{thm:lens} and Figure~\ref{fig:lens}). At a given level of industry profit, consumers are worst off when captive and contested consumers are perfectly separated, as in the discrimination regime of \citet{armstrong2019}. These outcomes lie on the straight line between Bertrand competition and monopoly. Consumers are best off in an unsegmented market in which the two firms have equally many captives. Every point in between is produced by an explicit market split into at most three segments.

Each edge is reached in only one way (Theorem~\ref{thm:edges}). Only unsegmented symmetric markets attain the upper edge. When the firms' captive bases differ and some consumers are contested, no segmentation reaches the upper edge: whatever information is released, consumers get strictly less than in a symmetric market with the same industry profit. The lower edge is attained only by segmentations that separate captive from contested consumers, such as perfect segmentation, and every market reaches it at exactly one point.

Except at Bertrand competition and monopoly, the upper edge itself lies strictly below the outcome in which every consumer pays one deterministic price and firms earn the same profit. No composition or information policy closes this gap, because it comes from the price dispersion that competition creates whenever some consumers are captive and some are contested. Under unit demand the gap vanishes and the lens collapses to a line.

\paragraph{Mechanism} Following \citet{armstrong2019}, I describe each price by the profit it earns (an \emph{offer}) and view a market as a lottery over the offer a randomly drawn consumer pays. Producer surplus is the mean of this lottery. Consumer surplus is the mean of a concave function of it, because at low prices a dollar of profit is simply a dollar transferred from consumers, while near the monopoly price it comes with consumers paying more and buying less, so it costs them more than a dollar. Because of this concavity, among markets with the same industry profit, consumers do worse the more dispersed the offers are. The shape of the lens therefore comes down to comparing dispersion across markets, and two comparisons do the work. Both appear in the proof of Proposition~4 of \citet{armstrong2019}. First, compare a market in which only one firm has captives with a market in which the two firms have equally many captives and industry profit is the same. In the first market, the firm with captives often gives up on the contested consumers and charges the monopoly price, while its rival, expecting this, undercuts more aggressively. As a result, prices are both higher and lower around the same mean: offers are a mean-preserving spread of those in the second market. Second, a mixture of symmetric markets is a mean-preserving spread of the single symmetric market with the same profit. I restate both in the strict form I need (Lemmas~\ref{lem:mps} and~\ref{lem:sym}).

My contribution is to combine these comparisons with the decomposition of \citet{csz2026}, which writes every market as a mixture of the two kinds above, into a complete description of the outcomes that segmentation can produce when the composition is unknown, including which markets attain each edge. For linear demand I also derive both edges in closed form, which measures how much consumers lose to price dispersion and how much to asymmetry and segmentation.

\paragraph{Outline} Section~\ref{sec:model} presents the model and the results from earlier work that I use. Section~\ref{sec:lens} characterizes the set of achievable outcomes, Section~\ref{sec:edges} identifies the markets that reach its edges, and Section~\ref{sec:example} works through the linear demand $D(p)=2-p$. Section~\ref{sec:conclusion} concludes. The longer proofs are in the Appendix.

\subsection{Related Literature}

The model of captive and contested consumers goes back to \citet{varian1980} and \citet{narasimhan1988}, who assume unit demand. \citet{armstrong2019} allow downward-sloping demand and study discrimination against captives. On the information side, \citet{bbm2015} characterize all combinations of consumer and producer surplus that segmentation can produce for a monopolist. With unit demand, \citet{bbm2020} and \citet{albrecht2020} show that nested segmentation maximizes producer surplus in the captive-consumer model. My closest antecedents are \citet{armstrong2019} and \citet{csz2026}, whose model, notation, and equilibrium characterization I use. All of these papers hold the composition of the market fixed. Letting it vary replaces a family of optimization problems with a single set whose shape is easy to describe. \citet{tekdir2026unknown} asks the same question for a monopolist whose distribution of values is unknown.

%=====================================================================
\needspace{5\baselineskip}
\section{Model}\label{sec:model}

\subsection{Consumers, Firms, and Markets}

The model is that of \citet{armstrong2019} in the notation of \citet{csz2026}. Two firms produce a homogeneous good at zero marginal cost and compete in prices. A unit mass of consumers have quasilinear preferences and a common demand curve $D(p)$. There are three kinds of consumer: \emph{captives of firm~1}, who consider only firm~1; \emph{captives of firm~2}; and \emph{contested} consumers, who buy from whichever firm charges less.

A \emph{segment} is a vector $q=(q_1,q_2)$ in
\[
\Q=\{(q_1,q_2)\in[0,1]^2: q_1+q_2\le 1\},
\]
where $q_i$ is the share of consumers captive to firm $i$ and $1-q_1-q_2$ is the share contested. The whole market before any segmentation is the \emph{prior} $\gamma=(\gamma_1,\gamma_2)\in\Q$. Two kinds of segment play a special role. A segment is \emph{symmetric} if $q_1=q_2$, so that both firms have the same number of captives. It is \emph{nested} if one coordinate is zero, say $(q,0)$, so that only one firm has captives.

\subsection{Measuring Prices by the Profit They Earn}

Let $\pi(p)=pD(p)$ be a firm's profit from one buying consumer at price $p$. It is maximized at the monopoly price $p^*$, where it equals $\ps$. Note that $\ps$ is profit \emph{per consumer}.

\begin{assumption}[\citealp{armstrong2019,csz2026}]\label{ass:elas}
$D$ is twice differentiable and strictly decreasing where positive, and the elasticity $\eta(p)=-pD'(p)/D(p)$ is strictly increasing and equals one at some price $p^*$.
\end{assumption}

Under Assumption~\ref{ass:elas}, $\pi(p)$ is strictly increasing on $[0,p^*]$, and no firm charges more than $p^*$ in equilibrium. As in \citet{armstrong2019}, it is convenient to describe each price by the profit it earns, $\pi\in[0,\ps]$. I call $\pi$ an \emph{offer} and write $p(\pi)$ for the price that earns it. A consumer who faces offer $\pi$ pays $p(\pi)$, gives the firm profit $\pi$, and keeps consumer surplus
\[
v(\pi)=\int_{p(\pi)}^{\infty}D(x)\,dx .
\]

\paragraph{Consumer surplus is concave in profit} \citet{armstrong2019} show that Assumption~\ref{ass:elas} makes $v$ strictly decreasing and strictly concave on $[0,\ps]$, and \citet{csz2026} use the same property. I recall the argument, because every result below uses it. Differentiating,
\[
v'(\pi)=-D(p)\,p'(\pi)=-\frac{D(p)}{\pi'(p)}=-\frac{1}{1-\eta(p)},\qquad p=p(\pi).
\]
At low prices demand is inelastic ($\eta\approx0$), so $v'\approx-1$ and an extra dollar of profit is simply a dollar taken from consumers. As the price approaches $p^*$, $\eta\to1$ and $v'\to-\infty$: the last dollars of profit require price increases that also destroy sales, and consumers lose much more than the firm gains. Since $\eta$ rises with the price, $v'$ falls, which is concavity. Consumers therefore prefer a sure offer to any lottery over offers with the same mean. With \emph{unit demand}, by contrast, every consumer buys one unit as long as the price is below a common valuation, prices are pure transfers, and $v$ is linear.

\subsection{Equilibrium in a Segment}

If $q_1+q_2=1$ there are no contested consumers, and each firm charges its captives the monopoly price. If $q=(0,0)$ every consumer is contested and Bertrand competition drives the price to zero. In all other segments, label the firms so that $q_1\ge q_2$ and define firm~1's \emph{captive-to-reach ratio}, as in \citet{armstrong2019},
\[
\rho=\frac{q_1}{1-q_2}\in(0,1).
\]
It is the share of firm~1's potential customers (its captives plus the contested consumers) who are captive.

The equilibrium of a segment is standard. \citet{narasimhan1988} derives it for a duopoly with unit demand in which each firm has loyal customers and the rest buy from the cheaper firm, and \citet{armstrong2019} use it for their uniform-pricing regime. I state it in the form of \citet{csz2026}.

\begin{lemma}[\citealp{narasimhan1988}; \citealp{armstrong2019}, Lemma~1; \citealp{csz2026}, Lemma~1]\label{lem:eq}
Let $q_1\ge q_2$, $q_1>0$ and $q_1+q_2<1$. The segment has a unique equilibrium. Both firms randomize over offers in $[\rho\ps,\ps]$ with distributions
\begin{equation}\label{eq:F}
F_1(\pi)=\frac{1-q_1}{1-q_1-q_2}\Bigl(1-\frac{\rho\ps}{\pi}\Bigr)\ \ (\pi<\ps),\qquad
F_2(\pi)=\frac{1-q_2}{1-q_1-q_2}\Bigl(1-\frac{\rho\ps}{\pi}\Bigr),
\end{equation}
and firm~1 places the remaining probability $\frac{q_1-q_2}{1-q_2}$ on the monopoly offer $\ps$. Firm~1 earns $q_1\ps$ and firm~2 earns $(1-q_1)\rho\ps$.
\end{lemma}

Each part of the lemma has a direct interpretation. Firm~1 can always earn $q_1\ps$ by ignoring the contested consumers and charging its captives the monopoly price, so it never accepts less. At offer $\pi$ it reaches at most its captives and the contested consumers, a mass $1-q_2$, so it never offers less than $\rho\ps$, the offer at which $(1-q_2)\pi=q_1\ps$. A firm that randomizes must be indifferent across its offers. Firm~1's indifference condition,
\[
q_1\pi+(1-q_1-q_2)\,\pi\,\bigl(1-F_2(\pi)\bigr)=q_1\ps,
\]
pins down firm~2's distribution, and firm~2's indifference condition pins down firm~1's. Because firm~1 has more captives, it has more to lose by competing, and with probability $\frac{q_1-q_2}{1-q_2}$ it gives up on the contested consumers and charges the monopoly price. Industry profit is
\begin{equation}\label{eq:P}
P(q)=q_1\ps+(1-q_1)\rho\ps .
\end{equation}

\subsection{A Market as a Lottery over Offers}

Consider a consumer drawn at random from segment $q$. A captive of firm $i$ pays firm $i$'s offer. A contested consumer pays the lower of two independent offers, whose distribution is $F_1+F_2-F_1F_2$. Averaging over the three kinds of consumer, the offer this consumer pays has distribution
\[
G_q=q_1F_1+q_2F_2+(1-q_1-q_2)(F_1+F_2-F_1F_2).
\]
Since every consumer transfers $\pi$ and keeps $v(\pi)$, producer and consumer surplus per consumer are
\begin{equation}\label{eq:welfare}
P(q)=\E_{G_q}[\pi],\qquad C(q)=\E_{G_q}[v(\pi)].
\end{equation}
\citet{armstrong2019} (their equation (15)) and \citet{csz2026} (their equations (3)--(5)) use the same representation. Producers receive the mean of the lottery $G_q$, and consumers receive the mean of a concave function of it. For the degenerate segments, $G_{(0,0)}=\delta_0$, and $G_q=\delta_\ps$ when $q_1+q_2=1$.

Substituting \eqref{eq:F} and writing $\xi=(\ps/\pi)^2$ gives the two distributions used below. For the symmetric segment $(s,s)$,
\begin{equation}\label{eq:Gsym}
G_{(s,s)}(\pi)=1-\frac{s^2(\xi-1)}{1-2s}\quad\text{on }\Bigl[\tfrac{s}{1-s}\ps,\ps\Bigr],\qquad P(s,s)=2s\ps,
\end{equation}
and for the nested segment $(q,0)$,
\begin{equation}\label{eq:Gnest}
G_{(q,0)}(\pi)=1-q^2\xi\quad\text{on }[q\ps,\ps),\qquad P(q,0)=(2-q)q\ps,
\end{equation}
with an atom of size $q^2$ at $\ps$.
The symmetric segment has no atom, since both firms always compete. The nested segment puts probability $q^2$ on the monopoly offer. Both distributions are affine in $\xi$ on their supports, which makes them easy to compare.

\subsection{Reducing Every Segment to Two}

\begin{figure}[!htbp]
\centering
\includegraphics[width=0.68\linewidth]{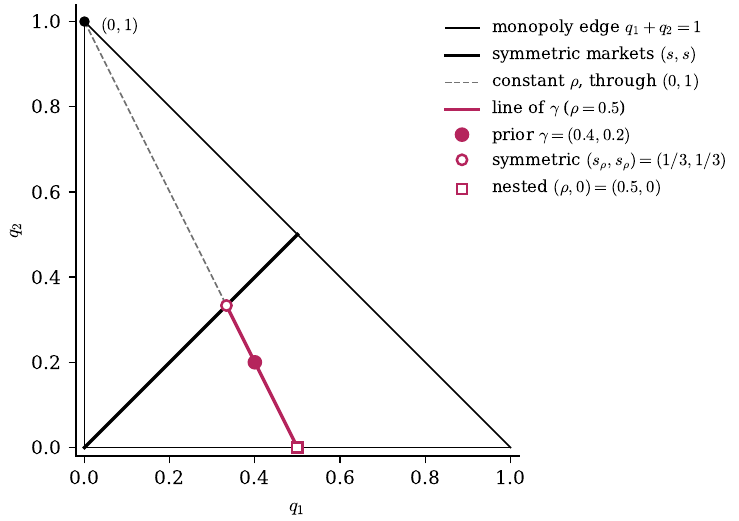}
\caption[The space of markets]{The space of markets $\Q$. Segments with a common captive-to-reach ratio $\rho$ lie on a straight line through the corner $(0,1)$, shown here for $\rho=0.5$. The market $\gamma=(0.4,0.2)$ lies on this line between the symmetric segment $(s_\rho,s_\rho)=(1/3,1/3)$ and the nested segment $(\rho,0)=(0.5,0)$; it is a mixture of the two, and its welfare is the same mixture (Lemma~\ref{lem:dec}).}
\label{fig:markets}
\end{figure}

Segments with a common captive-to-reach ratio $\rho$ lie on the line $q_1=\rho(1-q_2)$, which passes through the corner $(0,1)$ (Figure~\ref{fig:markets}). Where $q_1\ge q_2$, the line runs from the nested segment $(\rho,0)$ to the symmetric segment $(s_\rho,s_\rho)$, with $s_\rho=\rho/(1+\rho)$. Along such a line the offers have the same support $[\rho\ps,\ps]$, and the distribution of the offer a consumer pays moves linearly between those of the two endpoints. This gives the decomposition of \citet{csz2026}.

\begin{lemma}[\citealp{csz2026}, Lemma~2 and equation~(8)]\label{lem:dec}
Let $q_1\ge q_2$, $q_1>0$ and $q_1+q_2<1$. Then
\begin{equation}\label{eq:dec}
q=\lambda\,(s_\rho,s_\rho)+(1-\lambda)(\rho,0),\qquad \lambda=\frac{q_2(1+q_1-q_2)}{q_1}\in[0,1],
\end{equation}
and $(C(q),P(q))$ is the same $\lambda$-mixture of the outcomes of the two endpoints. Moreover, $\lambda<1$ whenever $q_1>q_2$.
\end{lemma}

I call $\lambda$ the \emph{symmetry index}. It equals one for symmetric segments and zero for nested ones. For welfare purposes every segment is a mixture of a symmetric and a nested segment with the same $\rho$, so it is enough to compare these two types.

\subsection{Segmentation and the Question}

Information is public. A \emph{segmentation} of $\gamma$ is a finitely supported distribution $\sigma$ over segments whose average is the prior, $\sum_q\sigma(q)\,q=\gamma$. Firms observe the segment and play its equilibrium. Two segmentations recur. \emph{Uniform pricing} is $\sigma=\delta_\gamma$. \emph{Perfect segmentation} puts weight $\gamma_1$ on $(1,0)$, weight $\gamma_2$ on $(0,1)$, and weight $1-\gamma_1-\gamma_2$ on $(0,0)$. It separates captive from contested consumers and is the discrimination regime of \citet{armstrong2019}.

The set of outcomes achievable in market $\gamma$ is
\[
W(\gamma)=\Bigl\{\textstyle\sum_q\sigma(q)\bigl(C(q),P(q)\bigr):\ \sigma\text{ is a segmentation of }\gamma\Bigr\},
\]
and the set of outcomes achievable in some market is $\U=\bigcup_{\gamma\in\Q}W(\gamma)$. \citet{csz2026} characterize the extreme points of $W(\gamma)$ in the directions of producer, consumer, and total surplus. I characterize $\U$. Two benchmarks frame the answer. The first is the curve
\[
\Gamma=\{(v(\pi),\pi):\pi\in[0,\ps]\}
\]
of outcomes in which every consumer pays the same sure offer. Its endpoints are Bertrand, $B=(v(0),0)$, and monopoly, $M=(v(\ps),\ps)$. The second is the straight line from $B$ to $M$,
\[
\ell(\pi)=v(0)-\bigl(v(0)-v(\ps)\bigr)\frac{\pi}{\ps}.
\]
Here and below, $\pi\in[0,\ps]$ also denotes a level of industry profit. An offer is the profit earned on a single consumer, and industry profit is the average offer, so the two are measured in the same units, and $v(\pi)$ is the consumer surplus of a sure offer that yields profit $\pi$. I keep $p$ for prices.

%=====================================================================
\needspace{5\baselineskip}
\section{The Welfare Lens}\label{sec:lens}

This section characterizes the set of achievable outcomes for every demand that satisfies Assumption~\ref{ass:elas}. Section~\ref{sec:example} illustrates each step with $D(p)=2-p$.

\subsection{Two Bounds from Concavity}

By \eqref{eq:welfare}, every segment outcome is $(\E_Gv,\E_G\pi)$ for some distribution $G$ on $[0,\ps]$. Since $v$ is concave, Jensen's inequality gives $C\le v(P)$, so no outcome lies to the right of $\Gamma$. A concave function lies above its chord, so $v\ge\ell$ on $[0,\ps]$ and hence $C=\E_Gv\ge\E_G\ell=\ell(P)$; no outcome lies to the left of the chord. The chord corresponds to the most dispersed lottery with a given mean (all mass on $0$ and $\ps$), and $\Gamma$ to the least dispersed (no randomness at all).

Where a market lands between these bounds depends on how dispersed its offers are. I compare dispersion in the sense of \citet{rothschild1970}. A distribution $F$ on $[0,\ps]$ is a \emph{mean-preserving spread} of $H$ if the two have the same mean and $\int_0^tF\ge\int_0^tH$ for all $t$. If so and $F\ne H$, then $\E_Fu<\E_Hu$ for every strictly concave $u$. A convenient sufficient condition, also used by \citet{armstrong2019}, is that the means agree and $F-H$ is nonnegative below some point and nonpositive above it, so that $F$ puts more mass on low values and correspondingly more on high values.

\subsection{Asymmetry Spreads Offers}

The next lemma is the first step in the proof of Proposition~4 of \citet{armstrong2019}, specialized to nested segments. I include the short proof because I need the strict inequality and the explicit distributions.

\begin{lemma}[asymmetry is a spread]\label{lem:mps}
For $q\in(0,1)$ let $s=q-q^2/2$. The nested segment $(q,0)$ and the symmetric segment $(s,s)$ yield the same profit, and $G_{(q,0)}$ is a mean-preserving spread of $G_{(s,s)}$. Hence $C(q,0)<C(s,s)$.
\end{lemma}

\begin{figure}[!tbp]
\centering
\includegraphics[width=0.8\linewidth]{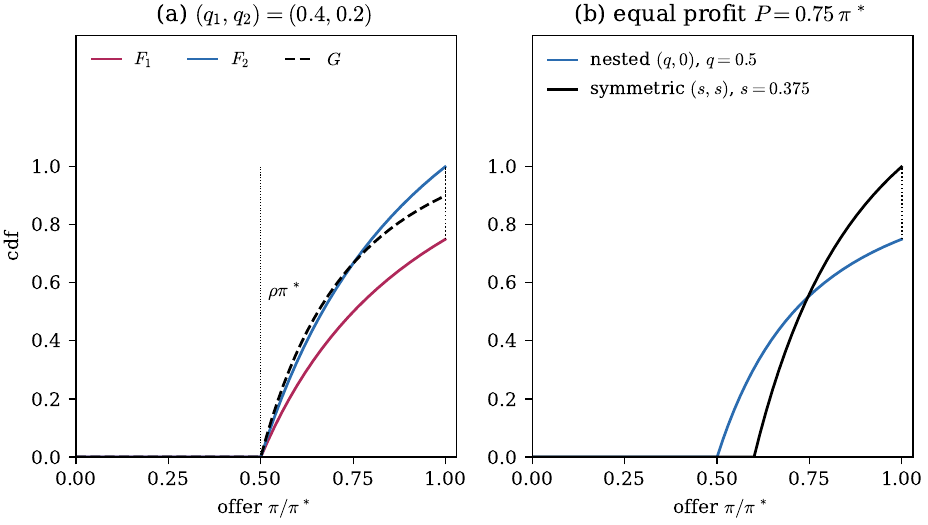}
\caption[Offer distributions in the symmetric, nested and an asymmetric segment]{Offer distributions in units of $\ps$. \textbf{(a)} Segment $(0.4,0.2)$: the two firms' distributions and $G$, the distribution of the offer a random consumer pays. \textbf{(b)} The nested segment $(0.5,0)$ and the symmetric segment $(0.375,0.375)$, which earn the same profit; the nested distribution is a mean-preserving spread of the symmetric one (Lemma~\ref{lem:mps}).}
\label{fig:offers}
\end{figure}

\paragraph{What Lemma~\ref{lem:mps} says} For any nested market there is a symmetric market that keeps industry profit the same but gives consumers higher surplus. The mean offer is the same in both, but consumers are worse off in the nested one, because its offers are more dispersed. Figure~\ref{fig:offers}(b) shows this for the nested segment $(0.5,0)$ and the symmetric segment $(0.375,0.375)$, which both earn $0.75\ps$. In the symmetric segment both firms compete for the contested consumers, and every offer lies in $[0.6\ps,\ps]$. In the nested segment the firm with captives often gives up on the contested consumers and charges the monopoly price, so a consumer pays the monopoly offer with probability $0.25$; its rival, expecting this, undercuts more, down to $0.5\ps$. The lemma is the reason an asymmetric market can never be best for consumers: in the proof of Theorem~\ref{thm:lens} it places every asymmetric segment below the symmetric curve, and in Theorem~\ref{thm:edges} it rules asymmetric segments out at the upper edge.

\begin{proof}
\emph{Equal profit.} Since $s=q-q^2/2=\tfrac12\bigl(1-(1-q)^2\bigr)<\tfrac12$, $(s,s)$ is a segment with some contested consumers. By \eqref{eq:Gsym}--\eqref{eq:Gnest}, $P(q,0)=(2-q)q\ps=2s\ps=P(s,s)$.

\emph{The nested distribution starts lower.} Write $G_N=G_{(q,0)}$ and $G_S=G_{(s,s)}$, whose supports start at $q\ps$ and at $\rho_S\ps$ with $\rho_S=s/(1-s)$. The inequality $\rho_S>q$ is equivalent to $s>q(1-s)$, which simplifies to $q^2(1-q)/2>0$.

\emph{One crossing.} Let $\Delta(\pi)=G_N(\pi)-G_S(\pi)$ be the gap between the two distribution functions, and recall that $\xi=(\ps/\pi)^2$ falls as $\pi$ rises. I show that $\Delta$ is nonnegative at low offers and negative at high ones.
\emph{Low offers.} Below $q\ps$ neither segment has offers, so $\Delta=0$. On $[q\ps,\rho_S\ps)$ only the nested segment has offers, so $G_S=0$ and $\Delta(\pi)=G_N(\pi)=1-q^2\xi\ge0$.
\emph{High offers.} On $[\rho_S\ps,\ps)$ both \eqref{eq:Gsym} and \eqref{eq:Gnest} apply, and
\[
\Delta(\pi)=\frac{s^2(\xi-1)}{1-2s}-q^2\xi .
\]
This is affine in $\xi$, so $\Delta$ is monotone on this range. At $\pi=\rho_S\ps$ the symmetric distribution is still zero, so $\Delta=1-q^2/\rho_S^2$, which is positive because $\rho_S>q$. As $\pi\to\ps$, $\xi\to1$, so $G_S\to1$ while $G_N\to1-q^2$, because the nested segment keeps an atom of size $q^2$ at the monopoly offer. Hence $\Delta\to-q^2<0$.
At $\pi=\ps$ both distributions equal one. So $\Delta$ changes sign once, from nonnegative to negative: compared with the symmetric segment, the nested segment puts more weight on low offers and on the monopoly offer.

\emph{Conclusion.} The means are equal, so $G_N$ is a mean-preserving spread of $G_S$, and the two differ. Since $v$ is strictly concave, $C(q,0)=\E_{G_N}v<\E_{G_S}v=C(s,s)$.
\end{proof}

\subsection{The Symmetric Frontier}

Let
\[
\Csym(\pi)=C\Bigl(\frac{\pi}{2\ps},\frac{\pi}{2\ps}\Bigr),\qquad \pi\in[0,\ps],
\]
be consumer surplus in the symmetric segment whose profit is $\pi$. It runs from $B$ (no captives) to $M$ (all consumers captive, half to each firm). \citet{csz2026}, Lemma~4, show that welfare is weakly concave along symmetric segments whenever the per-consumer objective is weakly concave, an argument they attribute to \citet{armstrong2019}, where it is the second step in the proof of Proposition~4. My uniqueness results need strict concavity.

\needspace{6\baselineskip}
\begin{lemma}[strict version of \citealp{csz2026}, Lemma~4]\label{lem:sym}
$\Csym$ is strictly concave on $[0,\ps]$.
\end{lemma}

The proof, in the Appendix, uses the same single-crossing argument as Lemma~\ref{lem:mps}: mixing two symmetric segments produces more dispersed offers than the single symmetric segment with the same average profit.

\paragraph{Why Lemma~\ref{lem:sym} is needed} The curve $\Csym$ will turn out to be the upper edge of the set of outcomes, and its shape matters twice. Concavity makes the set convex: splitting a symmetric market into two symmetric segments with different profits gives consumers no more than the single symmetric market with the average profit, so averaging segment outcomes, which is what every segmentation does, never rises above $\Csym$ (Theorem~\ref{thm:lens}). Strict concavity gives uniqueness (Theorem~\ref{thm:edges}(i)). If $\Csym$ were linear on some interval $[\pi_1,\pi_2]$, splitting a market into the symmetric segments with profits $\pi_1$ and $\pi_2$ would also reach the upper edge in between, and the unsegmented symmetric market would no longer be the only way to get there.

\subsection{Characterization}

\begin{theorem}[the welfare lens]\label{thm:lens}
Under Assumption~\ref{ass:elas},
\[
\U=\bigl\{(c,\pi):\ 0\le\pi\le\ps,\ \ \ell(\pi)\le c\le\Csym(\pi)\bigr\},
\]
and $\ell(\pi)<\Csym(\pi)<v(\pi)$ for every $0<\pi<\ps$. In particular, $\Gamma$ meets $\U$ only at $B$ and $M$.
\end{theorem}

At every profit level, then, consumer surplus can take any value between the chord and the symmetric curve, and no other; between $B$ and $M$ even the best case falls strictly short of a single sure price.

The proof is in the Appendix and has three steps. Step~1 shows that the two edges are apart: the symmetric segment with profit $\pi$ randomizes, so by Jensen's inequality $\Csym(\pi)$ lies strictly between $\ell(\pi)$ and $v(\pi)$. Step~2 shows that no outcome leaves the lens. Every segment is a mixture of a symmetric segment, whose outcome lies on the upper edge, and a nested segment, whose outcome Lemma~\ref{lem:mps} places strictly to the left of the upper edge. Concavity of $\Csym$ (Lemma~\ref{lem:sym}) keeps the mixture on or below the upper edge, and a segmentation only averages such outcomes. Step~3 shows that every point is reached. Fix a target $(c,\pi)$ and let $\alpha=(c-\ell(\pi))/(\Csym(\pi)-\ell(\pi))$ be its position on the horizontal slice of the lens at height $\pi$. Put weight $\alpha$ on the symmetric segment $(s,s)$ with $s=\pi/(2\ps)$, which reaches the upper edge, and split the remaining weight between the fully captive segment $(1,0)$ and the fully contested segment $(0,0)$ in the proportions $\pi/\ps$ and $1-\pi/\ps$, which together reach the lower edge. These segments average to the market
\begin{equation}\label{eq:construct}
\gamma=\frac{\pi}{\ps}\Bigl(1-\frac{\alpha}{2},\ \frac{\alpha}{2}\Bigr),
\end{equation}
whose segmentation into these three segments produces exactly $(c,\pi)$.

\paragraph{Two sources of consumer loss} For every achievable $(c,\pi)$ with $0<\pi<\ps$,
\begin{equation}\label{eq:shortfall}
v(\pi)-c=\underbrace{v(\pi)-\Csym(\pi)}_{>0}+\underbrace{\Csym(\pi)-c}_{\ge0}.
\end{equation}
The first term is the cost of competition itself: when some consumers are contested and some captive, firms must randomize, whatever the composition or the information. The second reflects asymmetry and segmentation. For $D(p)=2-p$ the first never exceeds about $0.018$ (near $\pi=0.76$), while the second, under perfect segmentation, reaches about $0.23$ near $\pi=0.75$, where the total shortfall $v-\ell$ peaks at $0.25$ (Section~\ref{sec:example}).

\paragraph{Unit demand} With unit demand at valuation $\ps$, $v$ is linear and \eqref{eq:welfare} gives $C(q)+P(q)=\ps$ in every segment. The lens collapses to the segment from $(\ps,0)$ to $(0,\ps)$: prices are pure transfers, so the gap between the edges comes entirely from the curvature of $v$. This is why the consumer analysis of \citet{csz2026} differs from the unit-demand models of \citet{bbm2020} and \citet{albrecht2020}.

%=====================================================================
\needspace{5\baselineskip}
\section{Who Reaches the Edges}\label{sec:edges}

The proof of Theorem~\ref{thm:lens} gives one way to reach each edge. There is no other, and this pins down what any given market can reach. Write $\bar\gamma=\gamma_1+\gamma_2$ for the total captive share.

\begin{theorem}[uniqueness at the edges]\label{thm:edges}
Under Assumption~\ref{ass:elas}:
\begin{enumerate}[label=(\roman*)]
\item \emph{Upper edge.} For $0<\pi<\ps$, a segmentation produces $(\Csym(\pi),\pi)$ only if it places all weight on the symmetric segment $\bigl(\frac{\pi}{2\ps},\frac{\pi}{2\ps}\bigr)$. Hence a market with $0<\bar\gamma<1$ reaches the upper edge if and only if $\gamma_1=\gamma_2$, and then only under uniform pricing.
\item \emph{Lower edge.} A segmentation produces a point of the lower edge only if every segment is fully contested, $q=(0,0)$, or fully captive, $q_1+q_2=1$. Hence a market with $0<\bar\gamma<1$ reaches the lower edge at exactly one point, $\bigl(\ell(\bar\gamma\ps),\bar\gamma\ps\bigr)$, the outcome of perfect segmentation.
\end{enumerate}
\end{theorem}
\begin{proof}
(i) Suppose segments $q^i$ with weights $w_i>0$ and outcomes $(c_i,\pi_i)$ produce $(\Csym(\pi),\pi)$. By Theorem~\ref{thm:lens} and concavity,
\[
\Csym(\pi)=\textstyle\sum_iw_ic_i\le\sum_iw_i\Csym(\pi_i)\le\Csym\bigl(\sum_iw_i\pi_i\bigr)=\Csym(\pi),
\]
so both inequalities bind. Strict concavity (Lemma~\ref{lem:sym}) forces $\pi_i=\pi$, and then $c_i=\Csym(\pi)$, for every $i$. A segment with $0<\pi_i<\ps$ is neither $(0,0)$ nor fully captive. If it were asymmetric, Lemma~\ref{lem:dec} with $\lambda<1$ and Lemma~\ref{lem:mps} would place it strictly below $\Csym$. So every $q^i$ is the symmetric segment with profit $\pi$.

For the second claim, note that uniform pricing in the symmetric segment $(s,s)$ produces the point $\bigl(\Csym(2s\ps),\,2s\ps\bigr)$. Conversely, if a segmentation of $\gamma$ reaches the upper edge at a profit $\pi\in(0,\ps)$, all its segments equal one symmetric segment, which must then be $\gamma$ itself. At $\pi=0$, every segment has $P(q)=0$, which by \eqref{eq:P} forces $q=(0,0)$ and so $\bar\gamma=0$. At $\pi=\ps$, every segment has $(1-q_1)(1-\rho)=0$, so $q_1+q_2=1$ and $\bar\gamma=1$.

(ii) If $\sum_iw_iC(q^i)=\ell\bigl(\sum_iw_iP(q^i)\bigr)$, then, because $C(q^i)\ge\ell(P(q^i))$ for each segment and $\ell$ is affine, $C(q^i)=\ell(P(q^i))$ for every segment. Since $v>\ell$ on $(0,\ps)$, this requires $G_{q^i}$ to put all its mass on $\{0,\ps\}$. A segment with $q_1+q_2<1$ other than $(0,0)$ has offers in $(\rho\ps,\ps)$ with positive probability, so it is excluded. Fully contested segments yield $B$ and fully captive ones yield $M$. For the weights to average to $\gamma$, the fully captive segments must carry total weight $\bar\gamma$, which gives the stated point.
\end{proof}

\begin{figure}[!htbp]
\centering
\includegraphics[width=0.72\linewidth]{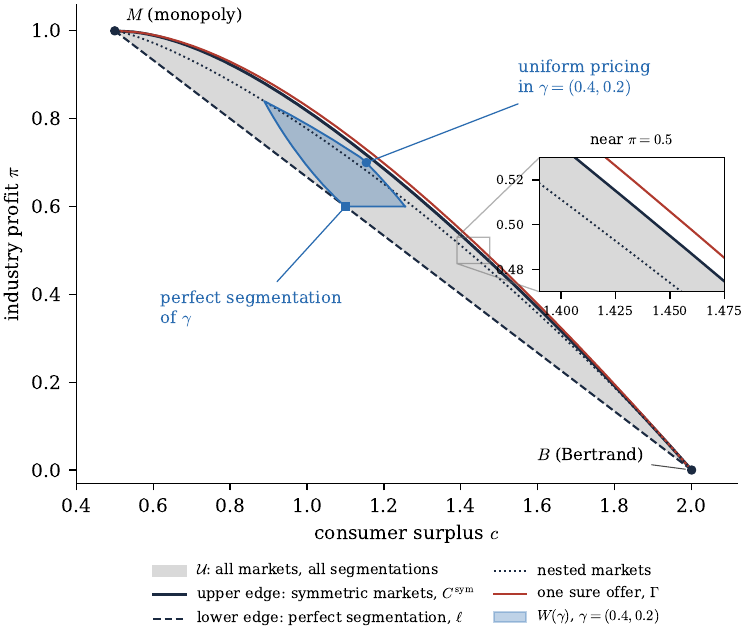}
\caption[The welfare lens for $D(p)=2-p$]{The welfare lens for $D(p)=2-p$ (and, in units of $\ps$, for every linear demand; see Section~\ref{sec:example}). Grey: $\U$, the outcomes of all markets and all segmentations. Upper edge: symmetric markets, $\Csym$. Lower edge: perfect segmentation, the chord $\ell$ from Bertrand ($B$) to monopoly ($M$). Red: $\Gamma$, one sure offer, which touches $\U$ only at $B$ and $M$. Dotted: nested markets. Blue: $W(\gamma)$, the outcomes of all segmentations of $\gamma=(0.4,0.2)$, with uniform pricing and perfect segmentation marked; it touches the chord once and stays strictly below the upper edge (Theorem~\ref{thm:edges}). Inset: the three curves near $\pi=0.5$.}
\label{fig:lens}
\end{figure}

Part~(i) is a welfare cost of asymmetry that holds for every information structure. If the firms have unequal captive bases and some consumers are contested, no segmentation raises consumer surplus to the level that a symmetric market with the same profit attains on its own. Segmentation can close part of this gap, as the consumer-optimal segmentations of \citet{csz2026} do by carving out a symmetric segment, but never all of it. Part~(i) is related to Proposition~4 of \citet{armstrong2019}, which shows that in a symmetric market uniform pricing gives consumers weakly more surplus, and firms weakly less profit, than any segmentation. Part~(ii) says that, in any market, discrimination against captives in the sense of \citet{armstrong2019} gives consumers strictly less than any segmentation with the same profit that uses a segment with both captive and contested consumers. Figure~\ref{fig:lens} illustrates both parts for $\gamma=(0.4,0.2)$.

\section{The Linear Demand \texorpdfstring{$D(p)=2-p$}{D(p)=2-p}}\label{sec:example}

This section works through one demand curve in full. It gives a concrete form to every object in Theorems~\ref{thm:lens} and~\ref{thm:edges} and derives closed forms for the two curves that bound the lens. \citet{armstrong2019} use the same demand curve to illustrate their results. Every number reported below is exact or follows from the closed forms in Proposition~\ref{prop:example}.

\subsection{Benchmarks}

With $D(p)=2-p$, profit per buying consumer is $\pi(p)=p(2-p)$, so the monopoly price is $p^*=1$ and $\ps=1$; offers and profits are measured in the same units. The elasticity $\eta(p)=p/(2-p)$ is increasing, so Assumption~\ref{ass:elas} holds on the relevant range $[0,p^*]$. (The kink of demand at $p=2$ lies above $p^*$ and plays no role.) Solving $p(2-p)=\pi$ for the lower root gives the price that earns offer $\pi$, and consumer surplus is the triangle under the demand curve:
\begin{equation}\label{eq:vex}
\begin{gathered}
p(\pi)=1-\sqrt{1-\pi},\\
v(\pi)=\frac{\bigl(2-p(\pi)\bigr)^2}{2}=\frac{\bigl(1+\sqrt{1-\pi}\bigr)^2}{2}=1-\frac{\pi}{2}+\sqrt{1-\pi}.
\end{gathered}
\end{equation}
Bertrand competition gives $B=(v(0),0)=(2,0)$, monopoly gives $M=(v(1),1)=(0.5,1)$, and the chord between them is $\ell(\pi)=2-1.5\pi$.

The last expression in \eqref{eq:vex} gives an exact measure of the role of dispersion. Taking expectations under the offer distribution $G$ of any segment,
\begin{equation}\label{eq:jensen}
C=1-\frac P2+\E_G\sqrt{1-\pi},\qquad v(P)-C=\sqrt{1-P}-\E_G\sqrt{1-\pi}.
\end{equation}
At a given profit, consumers rank markets by the expectation of the concave function $\sqrt{1-\pi}$, and their shortfall from a single sure price is exactly the Jensen gap of that function. Since $\sqrt{1-\pi}=1-p(\pi)$, the first equation can also be written as $C=2-P/2-\E_G[p]$: at a given profit, consumer surplus falls one for one with the mean price paid.

\paragraph{Other linear demands} The intercept and the slope only change units. For $D(p)=a-bp$, $\ps=a^2/(4b)$ and $v(\pi)=\ps-\frac\pi2+\ps\sqrt{1-\pi/\ps}$, so $v/\ps$ depends on $\pi$ only through $\pi/\ps$; by \eqref{eq:F}, so does every offer distribution. Measured in units of $\ps$, every outcome, every set $W(\gamma)$, and the lens are the same as for $D(p)=2-p$. Surplus and profit scale with $\ps$ and prices with $p^*=a/(2b)$; in particular, $C=2\ps-P/2-\tfrac a2\,\E_G[p]$, so the one-for-one link between consumer surplus and the mean price noted above is specific to $a=2$.

\subsection{Closed Forms for the Edges}

\begin{proposition}[closed forms for $D(p)=2-p$]\label{prop:example}
For $\pi\in(0,1)$, let $s=\pi/2$ and $T=\sqrt{(1-2s)/(1-s)}$. Then
\[
\Csym(\pi)=1-s+\frac{(1-s)(2-3s)\,T-s^2\atanh T}{2(1-2s)}.
\]
For the nested segment $(q,0)$ with $q\in(0,1)$, whose profit is $(2-q)q$, let $T=\sqrt{1-q}$. Then
\[
C(q,0)=1-\frac{(2-q)q}{2}+\Bigl(1-\frac q2\Bigr)T-\frac{q^2}{2}\atanh T.
\]
\end{proposition}

The proof, in the Appendix, integrates $\sqrt{1-\pi}$ against the distributions \eqref{eq:Gsym} and \eqref{eq:Gnest} and uses \eqref{eq:jensen}. The lower edge needs no computation: it is the chord $\ell(\pi)=2-1.5\pi$.

\subsection{Four Ways to Earn the Same Profit}

Fix industry profit at $\pi=0.5$ and compare four ways of producing it. Table~\ref{tab:example} collects the results.
\begin{enumerate}
\item \emph{One sure offer.} Every consumer pays $p(0.5)=1-\sqrt{0.5}\approx0.293$, and consumer surplus is $v(0.5)=(3+2\sqrt2)/4\approx1.457$. This point lies on $\Gamma$. No segment with contested consumers produces it, because such segments always have random prices.
\item \emph{Symmetric segment.} By \eqref{eq:Gsym}, $P(s,s)=2s$, so $s=0.25$: a quarter of consumers are captive to each firm and half are contested. Then $\rho=1/3$, both firms draw offers from $[1/3,1]$, which correspond to prices in $[0.184,1]$, and the offer a random consumer pays has distribution $G(\pi)=1-\tfrac18(\pi^{-2}-1)$. Proposition~\ref{prop:example} with $s=\tfrac14$ gives $T=\sqrt{2/3}$ and $\atanh T=\ln(\sqrt3+\sqrt2)$, so
\[
\Csym(0.5)=\tfrac34+\tfrac{15}{16}\sqrt{2/3}-\tfrac1{16}\ln\bigl(\sqrt3+\sqrt2\bigr)\approx1.444 .
\]
\item \emph{Nested segment.} By \eqref{eq:Gnest}, $(2-q)q=0.5$ gives $q=1-1/\sqrt2\approx0.293$, and only firm~1 has captives. Offers lie in $[0.293,1]$, which correspond to prices in $[0.159,1]$, and with probability $q^2=\tfrac32-\sqrt2\approx0.086$ the consumer pays the monopoly price $p^*=1$. Proposition~\ref{prop:example} gives $C(q,0)\approx1.415$.
\item \emph{Perfect segmentation.} Half of consumers are in fully captive segments (offer $1$) and half in fully contested segments (offer $0$). Consumer surplus is $\tfrac12v(1)+\tfrac12v(0)=1.25=\ell(0.5)$.
\end{enumerate}

\begin{table}[!htbp]
\centering
\small
\begin{tabular}{@{}llccc@{}}
\toprule
 & Market & Prices paid & Mean price & CS\\
\midrule
Sure offer & --- & $0.293$ & $0.293$ & $1.457$\\
Symmetric & $(0.25,0.25)$ & $[0.184,1]$ & $0.306$ & $1.444$\\
Nested & $(0.293,0)$ & $[0.159,1]^{\dagger}$ & $0.335$ & $1.415$\\
Perfect segm. & $\gamma_1+\gamma_2=0.5$ & $0$ or $1$ & $0.500$ & $1.250$\\
\bottomrule
\end{tabular}
\caption[Four ways to produce industry profit $\pi=0.5$ when $D(p)=2-p$]{Four ways to produce industry profit $\pi=0.5$ when $D(p)=2-p$. CS is consumer surplus. $^{\dagger}$With an atom at the monopoly price $1$. Profit is the same in every row. Consumer surplus falls as prices become more dispersed, and in every row consumer surplus plus the mean price equals $2-\pi/2=1.75$.}
\label{tab:example}
\end{table}

The ordering $1.457>1.444>1.415>1.250$ summarizes the main results. Each row adds dispersion to the price consumers pay while profit stays fixed: the first gap is the cost of competition (Theorem~\ref{thm:lens}), which no segmentation avoids, the second is the cost of asymmetry (Lemma~\ref{lem:mps}), and the third is the cost of separating captive from contested consumers. The mean-price column shows why: profit $p(2-p)$ is concave in the price, so at the same expected profit a lottery over prices has a higher average price than the sure price, and by \eqref{eq:jensen} consumers lose exactly that difference.

\subsection{A Given Market and a Given Point}

Take the prior $\gamma=(0.4,0.2)$. Its captive-to-reach ratio is $\rho=0.4/0.8=0.5$, and Lemma~\ref{lem:dec} writes it as $\gamma=0.6\,(\tfrac13,\tfrac13)+0.4\,(0.5,0)$, with symmetry index $\lambda=0.6$. Under uniform pricing, profit is $0.6\cdot\tfrac23+0.4\cdot0.75=0.7$, and by Proposition~\ref{prop:example} consumer surplus is $0.6\,\Csym(\tfrac23)+0.4\,C(0.5,0)=0.6\cdot1.227+0.4\cdot1.045\approx1.154$. Under perfect segmentation, profit falls to $\gamma_1+\gamma_2=0.6$ and consumer surplus to $\ell(0.6)=1.1$. This is the only point at which $W(\gamma)$ touches the lower edge (Theorem~\ref{thm:edges}(ii)).

Conversely, take a target point: profit $0.75$ and consumer surplus halfway between the edges, $c=\tfrac12(0.875+1.107)\approx0.991$. The construction in the proof of Theorem~\ref{thm:lens} gives $\alpha=\tfrac12$ and the market $\gamma=(0.5625,0.1875)$, split into the symmetric segment $(0.375,0.375)$, the fully captive segment $(1,0)$ and the fully contested segment $(0,0)$ with weights $\tfrac12$, $\tfrac38$ and $\tfrac18$.

%=====================================================================
\FloatBarrier
\needspace{5\baselineskip}
\section{Conclusion}\label{sec:conclusion}

When firms can tell captive consumers from contested ones, what consumers get depends on two things an outside observer rarely sees: how many consumers are captive to each firm, and how the market is split. I treat both as unknown and describe every pair of consumer surplus and industry profit they can produce. With downward-sloping demand these pairs form a lens. Its lower edge is the line between Bertrand competition and monopoly, and only segmentations that separate captive from contested consumers reach it. Its upper edge comes from unsegmented markets in which the two firms have equally many captives, and away from its endpoints even this edge falls short of a sure price with the same profit.

What separates the two edges is price dispersion. A firm with captives randomizes its price, and because consumer surplus is concave in profit, this randomization costs consumers. Unequal captive bases spread prices further, and so does segmentation. For a given industry profit, consumers are best off when the two firms have equally many captives and the market is not segmented, and worst off when captive and contested consumers are fully separated. With unit demand the lens collapses to a line, and none of this matters. Natural next steps are general demand, more than two firms, and choosing an information policy when the composition is unknown.

\clearpage
\bibliographystyle{plainnat}
\bibliography{references}

\clearpage
\section*{Appendix: Proofs}\addcontentsline{toc}{section}{Appendix: Proofs}

\needspace{5\baselineskip}
\subsection*{Proof of Lemma~\ref{lem:sym}}\label{app:sym}

Write $G_s=G_{(s,s)}$ and $\rho_s=s/(1-s)$, which is increasing in $s$. By \eqref{eq:Gsym}, for $\pi\in(0,\ps)$,
\[
G_s(\pi)=\max\{0,h_s(\pi)\},\qquad h_s(\pi)=1-(\xi-1)\frac{s^2}{1-2s},
\]
and $G_0=\delta_0$ agrees with $h_0\equiv1$. Since $s\mapsto s^2/(1-2s)$ has second derivative $2/(1-2s)^3>0$ and $\xi>1$ for $\pi<\ps$, the map $s\mapsto h_s(\pi)$ is strictly concave on $[0,\frac12)$.

Fix $0\le s_L<s_H<\frac12$ and $\theta\in(0,1)$, and let $s=\theta s_L+(1-\theta)s_H$ and $\bar G=\theta G_{s_L}+(1-\theta)G_{s_H}$. Because $P(s,s)=2s\ps$ is linear in $s$, $\bar G$ and $G_s$ have the same mean. Using $\rho_{s_L}<\rho_s<\rho_{s_H}$, consider three ranges of offers:
\begin{itemize}
\item on $(0,\rho_s\ps)$, $G_s=0\le\bar G$;
\item on $[\rho_s\ps,\rho_{s_H}\ps)$, $\bar G-G_s=\theta h_{s_L}-h_s$ is affine in $\xi$. It is nonnegative at $\rho_s\ps$, where $h_s=0$, and at $\rho_{s_H}\ps$, where $h_{s_H}=0$, it equals $\theta h_{s_L}+(1-\theta)h_{s_H}-h_s<0$;
\item on $[\rho_{s_H}\ps,\ps)$, all three distributions equal their $h$, and strict concavity in $s$ gives $\bar G<G_s$.
\end{itemize}
So $\bar G-G_s$ is nonnegative and then negative, and $\bar G$ is a mean-preserving spread of $G_s$ that differs from it. Since $v$ is strictly concave,
\[
\theta\Csym(2s_L\ps)+(1-\theta)\Csym(2s_H\ps)=\E_{\bar G}v<\E_{G_s}v=\Csym(2s\ps),
\]
which is strict concavity on $[0,\ps)$. As $s\to\frac12$, $G_s$ converges weakly to $\delta_\ps$, so $\Csym$ is continuous at $\ps$ with value $v(\ps)$, and concavity extends to $[0,\ps]$. Strictness extends as well. If the chord inequality held with equality at an interior point of a chord ending at $\ps$, the concave function $\Csym$ would be affine on that chord, and hence on a subinterval of $[0,\ps)$, contradicting strict concavity there. \qed

\needspace{8\baselineskip}
\subsection*{Proof of Theorem~\ref{thm:lens}}\label{app:lens}

Call the right-hand side the lens, $L$. Its lower boundary $\ell$ is linear and its upper boundary $\Csym$ is concave (Lemma~\ref{lem:sym}), so $L$ is convex. I show that the lens has a gap at every interior profit level, that no market and no segmentation leaves it, and that every point in it is reached.

\emph{Step 1: the gap.} Fix $0<\pi<\ps$. The symmetric segment with profit $\pi$ has a nondegenerate offer distribution $G$ with mean $\pi$ and with mass on $(0,\ps)$. Strict Jensen gives $\Csym(\pi)=\E_Gv<v(\pi)$. Since $v$ is strictly concave, $v>\ell$ on $(0,\ps)$, and since $\ell$ is affine, $\Csym(\pi)=\E_Gv>\E_G\ell=\ell(\pi)$.

\emph{Step 2: no outcome leaves $L$.} I first show that the outcome of every single segment lies in $L$ and then that averaging over segments keeps it there.

\emph{Lower bound for a segment.} The profit $P(q)$ of a segment $q$ lies in $[0,\ps]$. Since $v$ is concave, it lies above the chord: $v\ge\ell$ on $[0,\ps]$. Because $\ell$ is affine, $C(q)=\E_{G_q}v\ge\E_{G_q}\ell=\ell(P(q))$.

\begin{figure}[!htbp]
\centering
\includegraphics[width=0.72\linewidth]{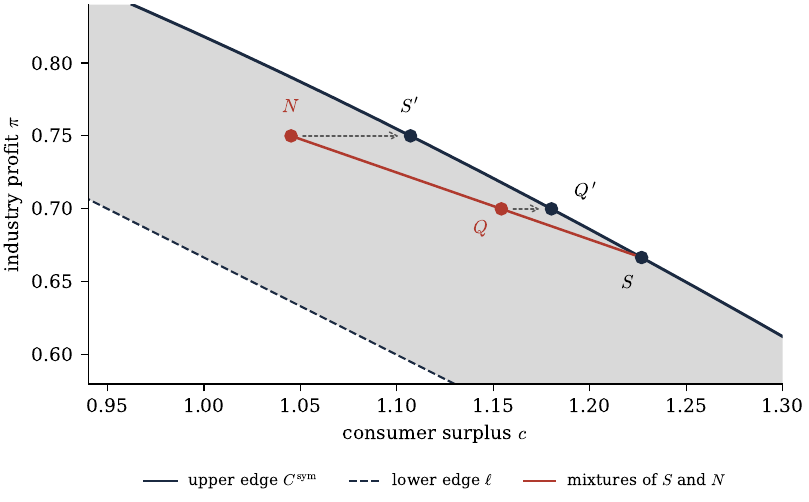}
\caption[Step 2 of the proof of Theorem 1]{Step 2 of the proof of Theorem~\ref{thm:lens} for $D(p)=2-p$ and the segment $q=(0.4,0.2)$. The outcome $Q$ of $q$ is a mixture of the outcomes of the symmetric segment $S$, on the upper edge, and the nested segment $N$. Lemma~\ref{lem:mps} moves $N$ to $S'$, the symmetric segment with the same profit, which accounts for almost all of the gap between $Q$ and $Q'$. Concavity of $\Csym$ (Lemma~\ref{lem:sym}) then lifts the average of $S$ and $S'$ the small remaining distance to $Q'$, the point of the upper edge at the same profit.}
\label{fig:step2}
\end{figure}

\emph{Upper bound for a segment.} The fully contested segment $(0,0)$ yields $B$ and fully captive segments yield $M$, and both points are on $\Csym$. Take any other segment $q$, labeled so that $q_1\ge q_2$, with captive-to-reach ratio $\rho=q_1/(1-q_2)$. Three segments enter the argument (Figure~\ref{fig:step2}):
\begin{itemize}
\item $S=(s_\rho,s_\rho)$, the symmetric segment with the same ratio $\rho$ as $q$;
\item $N=(\rho,0)$, the nested segment with the same ratio $\rho$;
\item $S'$, the symmetric segment that earns the same profit as $N$.
\end{itemize}
By Lemma~\ref{lem:dec}, $q=\lambda S+(1-\lambda)N$ for some $\lambda\in[0,1]$, and the outcome of $q$ is the same mixture of the outcomes of $S$ and $N$. Hence
\begin{align*}
C(q)&=\lambda\,C(S)+(1-\lambda)\,C(N)\\
&\le\lambda\,C(S)+(1-\lambda)\,C(S')\\
&=\lambda\,\Csym\bigl(P(S)\bigr)+(1-\lambda)\,\Csym\bigl(P(N)\bigr)\\
&\le\Csym\bigl(\lambda P(S)+(1-\lambda)P(N)\bigr)=\Csym\bigl(P(q)\bigr).
\end{align*}
The first inequality replaces $N$ by $S'$. The two segments earn the same profit, and by Lemma~\ref{lem:mps} the nested one gives consumers strictly less, so the inequality is strict whenever $\lambda<1$, that is, whenever $q_1>q_2$. The next equality holds because $S$ and $S'$ are symmetric, so their outcomes lie on $\Csym$. The second inequality is the concavity of $\Csym$ (Lemma~\ref{lem:sym}): two symmetric segments with different profits give consumers, on average, no more than the single symmetric segment with the average profit. The last equality uses $P(q)=\lambda P(S)+(1-\lambda)P(N)$, again from Lemma~\ref{lem:dec}. Weak concavity is enough here; the strict loss comes from Lemma~\ref{lem:mps}. For $D(p)=2-p$ and $q=(0.4,0.2)$, the three expressions in the chain are $1.154$, $1.179$ and $1.180$, all at profit $0.7$: almost all of the gap comes from replacing $N$ by $S'$.

\emph{Segmentations.} The outcome of a segmentation is the weighted average of the outcomes of its segments. Each of these lies in $L$, and $L$ is convex, so the average lies in $L$ as well.

\emph{Step 3: every point of $L$ is reached.} At $\pi=0$ and $\pi=\ps$ the lens is the single point $B$ or $M$, produced by the market $(0,0)$ or $(1,0)$. Fix $(c,\pi)\in L$ with $0<\pi<\ps$. The idea is to produce the two ends of the horizontal slice of the lens at height $\pi$ and then to mix them.
\begin{itemize}
\item The right end, $(\Csym(\pi),\pi)$, is produced by the symmetric segment $(s,s)$ with $s=\pi/(2\ps)$, priced without further segmentation.
\item The left end, $(\ell(\pi),\pi)$, is produced by perfect segmentation of a market whose total captive share is $\pi/\ps$. Its fully captive segments, with total weight $\pi/\ps$, each yield $M$, and its fully contested segments, with weight $1-\pi/\ps$, yield $B$. Average profit is $\pi$, and because $\ell$ is the straight line from $B$ to $M$, average consumer surplus is $\ell(\pi)$.
\end{itemize}
The target lies on the slice, so $c=\alpha\Csym(\pi)+(1-\alpha)\ell(\pi)$ with
\[
\alpha=\frac{c-\ell(\pi)}{\Csym(\pi)-\ell(\pi)}\in[0,1],
\]
which is well defined by Step~1. Combine the two pieces with weights $\alpha$ and $1-\alpha$: weight $\alpha$ on $(s,s)$, weight $(1-\alpha)\pi/\ps$ on the fully captive segment $(1,0)$, and weight $(1-\alpha)(1-\pi/\ps)$ on the fully contested segment $(0,0)$. These segments average to
\[
\gamma=\alpha\,(s,s)+(1-\alpha)\frac{\pi}{\ps}\,(1,0)+(1-\alpha)\Bigl(1-\frac{\pi}{\ps}\Bigr)(0,0)
=\Bigl(\alpha s+(1-\alpha)\frac{\pi}{\ps},\ \alpha s\Bigr).
\]
Substituting $s=\pi/(2\ps)$ gives
\[
\gamma=\Bigl(\Bigl(1-\frac{\alpha}{2}\Bigr)\frac{\pi}{\ps},\ \frac{\alpha}{2}\,\frac{\pi}{\ps}\Bigr)=\frac{\pi}{\ps}\Bigl(1-\frac{\alpha}{2},\ \frac{\alpha}{2}\Bigr),
\]
which is \eqref{eq:construct}.
The captive shares add up to $\gamma_1+\gamma_2=\pi/\ps$ whatever the value of $\alpha$: the total share of captive consumers is fixed by the target profit, and $\alpha$ only decides how it is divided between the two firms. Since $\pi/\ps\le1$, $\gamma$ lies in $\Q$, so the segments form a segmentation of the market $\gamma$. Its profit is $\alpha\pi+(1-\alpha)\pi=\pi$, and its consumer surplus is $\alpha\Csym(\pi)+(1-\alpha)\ell(\pi)=c$. \qed

\begin{figure}[!htbp]
\centering
\includegraphics[width=0.92\linewidth]{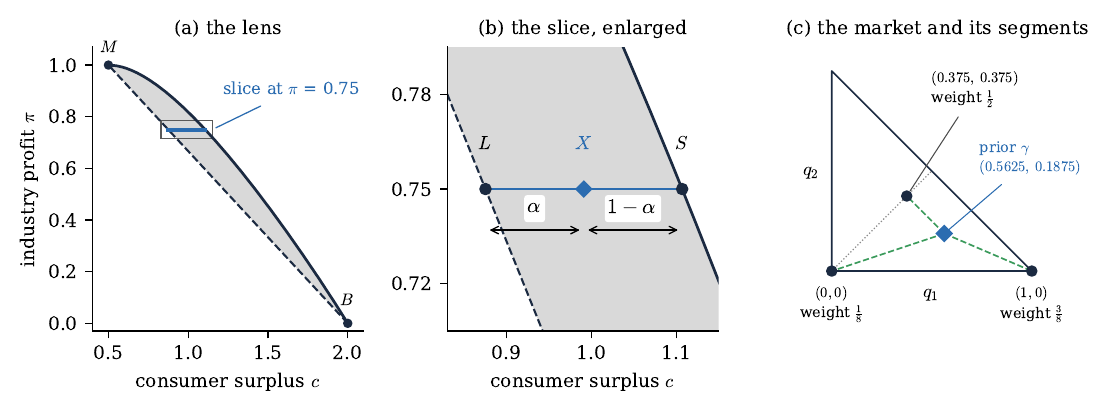}
\caption[Producing a chosen surplus pair]{Producing a chosen pair $(c,\pi)$ for $D(p)=2-p$, at profit $\pi=0.75$ and consumer surplus halfway across the lens ($\alpha=\frac12$). \textbf{(a)} The lens, with the slice at $\pi=0.75$; $B$ is the all-contested outcome and $M$ the all-captive outcome. \textbf{(b)} The slice, enlarged. $S\approx(1.107,0.75)$ is the outcome of the symmetric segment $(0.375,0.375)$ on the upper edge, $L=(0.875,0.75)$ is perfect segmentation on the chord, and the target $X\approx(0.991,0.75)$ lies a share $\alpha$ of the way from $L$ to $S$. \textbf{(c)} The same construction in the space of markets. The prior $\gamma=(0.5625,0.1875)$ is the average of the symmetric segment $(0.375,0.375)$ with weight $\frac12$, the fully captive segment $(1,0)$ with weight $\frac38$, and the fully contested segment $(0,0)$ with weight $\frac18$.}
\label{fig:construct}
\end{figure}

\needspace{10\baselineskip}
\paragraph{Reading the construction} Figure~\ref{fig:construct} draws the construction for $D(p)=2-p$. Moving the target along the slice changes consumer surplus but not profit, so in the figure the target moves horizontally: left as consumer surplus falls, right as it rises. By \eqref{eq:construct}, every market built this way has the same total captive share, $\pi/\ps$. The reason is that each of the three segments earns as much as it would if every captive consumer paid $\ps$ and every contested consumer paid nothing. The symmetric segment $(s,s)$ earns $P(s,s)=2s\ps$, the fully captive segment earns $\ps$, and the fully contested segment earns zero. Industry profit is therefore $\ps$ times the captive share, and fixing profit fixes the share. Only the arrangement of the captives changes. At the right end of the slice ($\alpha=1$), where consumer surplus is highest, the captives are split equally between the two firms and the market is left unsegmented. As the target moves left, $\alpha$ falls, the gap between the firms' captive shares grows, $\gamma_1-\gamma_2=(1-\alpha)\pi/\ps$, and more captives sit in segments of their own, away from contested consumers. At the left end ($\alpha=0$), where consumer surplus is lowest, firm~1 holds every captive and all of them are separated from the contested consumers. This is perfect segmentation.

\needspace{5\baselineskip}
\subsection*{Proof of Proposition~\ref{prop:example}}\label{app:closed}

Here $\ps=1$, and I write $x=\pi\in[0,1]$ for the offer. By \eqref{eq:jensen} it suffices to compute $E=\E_G\sqrt{1-x}$; an atom at $x=1$ contributes zero. By \eqref{eq:Gsym}--\eqref{eq:Gnest}, the continuous parts have densities
\[
g_{(s,s)}(x)=\frac{2s^2}{(1-2s)\,x^3}\ \text{ on }[\rho_s,1],\qquad g_{(q,0)}(x)=\frac{2q^2}{x^3}\ \text{ on }[q,1).
\]
Both reduce to $J(r)=\int_r^1\sqrt{1-x}\,x^{-3}dx$. The substitution $x=1-t^2$ gives $J(r)=\int_0^{T}2t^2(1-t^2)^{-3}dt$ with $T=\sqrt{1-r}$, and differentiation confirms
\begin{align*}
&\frac{d}{dt}\Bigl[\frac{t}{2(1-t^2)^2}-\frac{t}{4(1-t^2)}-\frac14\atanh t\Bigr]\\
&\qquad=\frac{\tfrac12(1+3t^2)-\tfrac14(1-t^4)-\tfrac14(1-t^2)^2}{(1-t^2)^3}
=\frac{2t^2}{(1-t^2)^3}.
\end{align*}
Since $1-T^2=r$, $J(r)=\frac{T}{2r^2}-\frac{T}{4r}-\frac14\atanh T$. For the symmetric segment, $r=\rho_s=s/(1-s)$, so $s^2/r^2=(1-s)^2$ and $s^2/r=s(1-s)$, and
\[
\begin{aligned}
E&=\frac{2s^2}{1-2s}J(\rho_s)=\frac{(1-s)^2T-\tfrac12s(1-s)T-\tfrac12s^2\atanh T}{1-2s}\\
&=\frac{(1-s)(2-3s)T-s^2\atanh T}{2(1-2s)}.
\end{aligned}
\]
For the nested segment, $E=2q^2J(q)=T-\frac q2T-\frac{q^2}{2}\atanh T$. Substituting into \eqref{eq:jensen} with $P=2s$ and $P=(2-q)q$ gives the formulas. \qed

\end{document}